\documentstyle[prb,aps,graphicx,psfrag]{revtex}

\begin{document}
\twocolumn[\hsize\textwidth\columnwidth\hsize\csname@twocolumnfalse\endcsname

\title{Electromechanics of charge shuttling in dissipative
  nanostructures} 
\author{T. Nord \and L.Y. Gorelik \and R.I.
  Shekhter, and M. Jonson} \address{Department of Applied Physics,
  Chalmers University of Technology \\ and G\"oteborg University,
  SE-412 96 G\"oteborg, Sweden} \date{\today} \maketitle

\begin{abstract}
  We investigate the current-voltage (IV) characteristics of a model
  single-electron transistor where mechanical motion, subject to
  strong dissipation, of a small metallic grain in tunneling contact
  with two electrodes is possible. The system is studied both by using
  Monte Carlo simulations and by using an analytical approach. We show
  that electromechanical coupling results in a highly nonlinear
  IV-curve. For voltages above the Coulomb blockade threshold, two
  distinct regimes of charge transfer occur: At low voltages the
  system behaves as a static, asymmetric double junction and tunneling
  is the dominating charge transfer mechanism. At higher voltages an
  abrupt transition to a new "shuttle" regime appears, where the grain
  performs an oscillatory motion back and forth between the
  electrodes. In this regime the current is mainly mediated by charges
  that are carried on the grain as it moves from one electrode to the
  other.
\end{abstract}

\pacs{PACS numbers: 73.23.HK, 85.35.GV, 85.85.+j}
%73.23.HK % Coulomb blockade; single-electron tunneling
%73.40.GK % Tunneling
%85.35.GV % Single electron devices
%85.85.+j % Micro- and nano-electromechanical systems (MEMS/NEMS) and
         % devices
]
\narrowtext

\section{Introduction}
The mechanical properties of mesoscopic conductors and their influence
on charge transport are very much in the focus of recent solid state
physics research. Certain anomalous behavior of
nanowires\cite{a:96_rubio,a:98_garcia} and the electrostatically
controlled deformation of carbon nanotubes\cite{a:00_rueckes} are
examples of an interplay between electrical and mechanical degrees of
freedom that appear on the nanometer length scale. Other examples
where the heteroelastic nature of a material crucially affects
single-electron tunneling have been found in studies of self-assembled
metal-organic composite structures.

The relevant scenario associated with a strong electromechanical
coupling is that significant deformations occur as a result of large
Coulomb forces acting on charge accumulated in some small region, for
instance, in a metallic cluster. Recently, a model system containing
such a coupling was considered by Gorelik {\em et
  al.}\cite{a:98_gorelik}, who proposed a single-electron tunneling
device containing a movable metallic cluster in tunneling contact with
bulk metallic electrodes. In this device mechanically soft organic
links serve both as elastic springs, keeping the cluster in place, and
as tunnel barriers with resistances that are exponentially sensitive
to the deformation of the springs. An important consequence of the
interplay between single electron tunneling and the mechanical
vibration of the cluster in this model is the electromechanical
instability predicted in Ref. \onlinecite{a:98_isacsson}: If a large
enough bias voltage is applied between the electrodes, the equilibrium
position of the grain looses its stability and cluster vibrations
develop. Such vibrations give rise to a new mechanism of charge
transfer, where electrons are transported through the system by the
metallic cluster which performs shuttle motion between the electrodes.
The electric current, $I=2Nef$, associated with this mechanism does
not depend on the tunnel transparencies, and is only determined by the
frequency, $f=\omega/2\pi$, of the elastic vibrations of the cluster
and the number, $N$, of electrons carried by the cluster. Experimental
evidence for a coupling between electron transfer and vibrational
degrees of freedom has been found both for
macroscopic\cite{a:99_tuominen} and
microscopic\cite{a:98_erbe,a:01_erbe,a:00_park} systems. Different
aspects of this phenomenon has also been theoretically investigated in
several
articles\cite{a:99_weiss,a:01_boese,a:01_nishiguchi,a:01_gorelik,a:01_isacsson,a:01_fedorets} 

In the work discussed above\cite{a:98_gorelik,a:98_isacsson} it was
shown that a large damping constant, $\gamma$, is detrimental for the
development of the shuttle instability and in the limit where
$\gamma\gtrsim f$, elastic shuttling of the charge becomes impossible.
The mechanical lability of the system, however, is still a dominating
feature of the charge transport even in the limit of strong
dissipation. What the consequences of such a lability would be is a
question which needs to be answered. This is not only an academic
question since coupling to intramolecular vibrations inside deformable
organic molecules carrying current as well as friction in the medium
through which a metallic cluster is moving may cause significant
dissipation. The dissipative limit of electromechanical mesoscopic
structures with movable parts is therefore important for understanding
the functioning of realistic nanometer size structures. Our objective
in the present work is to study this limit.

We will consider the model system illustrated in \mbox{Fig.
  \ref{fig:system}}. The current flow between the metallic electrodes
is due to the tunneling of electrons between the electrodes and the
metallic cluster. This is assisted by the displacement of the cluster.
An electrostatic force acts on the charged grain if a finite bias
voltage is applied between the electrodes. The one-dimensional
dynamics of the cluster is also governed by an elastic restoring force
and a friction force. In contrast to the approach developed in
\mbox{Ref. \onlinecite{a:98_isacsson}} we will consider the limit
where the electric force dominates over the elastic force, which means
that the dynamics of the charged cluster is determined by the
interplay between Coulomb forces and friction. This, however, does not
mean that the elastic forces can be totally neglected. For bias
voltages slightly above the Coulomb blockade threshold when the
cluster is in the center of the system, the dynamics of the cluster is
actually very sensitive to the value of the elastic force. The low
temperature non linear charge transport through the system is affected
both by the Coulomb blockade phenomenon and the mechanical motion of
the cluster. These two phenomena are coupled since the threshold
voltage for electron tunneling depends on the junction capacitances
which in turn depend on the cluster position with respect to the
electrodes. A general property is that the threshold voltage increases
when the distance between the cluster and an electrode decreases.

In order to understand qualitatively the electromechanical
charge-transport scenario, let us consider a neutral cluster situated
in its equilibrium position between the electrodes where the voltage
threshold for electron tunneling has a minimum value, $V_0$.  At zero
temperature no tunneling is possible for voltages $V<V_0$, where $V$
is the bias voltage applied between the electrodes. For $V>V_0$
tunneling onto the cluster becomes possible and the cluster can be
charged. It is easy to understand that the direction of motion of the
charged cluster, due to the Coulomb forces, will be away from the
electrode which has supplied the extra charge to the cluster. After
some time the extra charge will disappear, usually to the nearest
electrode, which makes the cluster charge zero again. An important
question at this stage is whether one more tunneling event to the
nearest electrode is possible or not. The answer is not evident since
the electrostatic threshold is different from the one at the initial
point in the middle of the system. As we will see, depending on the
applied bias voltage, we can have one of two possible situations. For
voltages \mbox{$V_0 < V < V_t$},where $V_t$ is a threshold voltage
which will be treated in more detail in \mbox{Sec.
  \ref{sec:an_desc_shuttle_reg}}, the extra tunneling event is not
possible. In this case the cluster is almost trapped near the
electrode. Small oscillations in the vicinity of the trapping point
are possible due to the action of the weak elastic force, but the
cluster will not be pushed back by Coulomb forces. In this case the
conductance is not assisted by significant cluster displacements
between the electrodes. We call this regime the \emph{tunneling
  regime} since the charge transfer is very similar to the
conventional single electron transport in a static system.

If $V>V_t$ there is a possibility for another tunneling event between
the grain and the nearest lead to happen after the extra charge has
tunneled off the cluster. This event changes the sign of the net
charge on the grain. In this case the cluster can be pushed by the
Coulomb force towards the more distant electrode where the above
described process repeats itself. The conductance is now assisted by
significant displacements of the grain and this scenario is
qualitatively similar to the shuttle vibrations in fully elastic
electromechanical structures\cite{a:98_gorelik}. We call this regime
the \emph{shuttle regime} of charge transport. A sharp transition,
corresponding to a current jump, occurs in a small voltage interval
between the two regimes.

\section{Model system}
\label{sec:model-system}
We will consider a model based on the picture in \mbox{Fig.
  \ref{fig:system}}. This is a simplified model which, however,
retains many of the interesting features of a ``real'' system.  The
system consists of a metallic grain of mass $M$ placed in the gap
between two bulk leads separated by a distance $L$. The displacement
of the grain from the center of the system is measured by the
coordinate $X$\footnote{Our approach is based on a classical
  description of the grain displacement and is different from
  approaches where quantum cluster vibration assisted tunneling to the
  grain is considered\cite{a:01_boese}}. We consider only 1D-motion
of the grain between the leads.  A bias voltage $V$ is applied between
the leads. In this simplified case we take into account only three
different forces acting on the grain; a linear elastic restoring force
$F_{el}=-kX$, a dissipative damping force $F_{diss}=-\gamma_d\dot{X}$,
and an electrostatic force $F_q$.  The electrostatic force is a
function of the bias voltage, $V$, and the charge, $Q$, on the grain:
\begin{equation}
  \label{eq:el_force_2}
  F_q = \frac{QV}{L} + \frac{X}{C_0L^2}Q^2 \,.
\end{equation}
Here $C_0$ is a capacitance constant determined by the geometry of the
system. To get this expression we assume that the capacitances $C_L$
between the left lead and the grain and $C_R$ between the grain and
the right lead can be approximated as parallel plate capacitors and
that all other capacitances can be neglected. The first term in
\mbox{Eq. (\ref{eq:el_force_2})} can be understood as the force from
an effective electrostatic field $V/L$ in the junction, which couples
to the extra charge on the grain. The second term can be thought of as
the interaction of the charge on the grain with image charges in the
two leads. Note that the last term in \mbox{Eq. (\ref{eq:el_force_2})}
always results in an attraction of the charged grain towards the
nearest lead. If we take these forces into account we can write the
equation of motion for the grain as:
\begin{equation}
  \label{eq:4}
  M\ddot{X} + \gamma_d\dot{X} + kX = \frac{QV}{L} + \frac{X}{C_0L^2}Q^2.
\end{equation}

We can now consider the Coulomb blockade regime where the Coulomb
charging energy, $E_c = e^2/2C$ is larger than both quantum and
thermal fluctuations, $E_c \gg \hbar/RC, \beta^{-1}$, where $R$ is the
smallest tunneling resistance possible in the system, and $\beta$ is
the inverse temperature. We thus assume that 
\begin{displaymath}
  R(X) \gg R_Q \equiv \frac{\pi\hbar}{2e^2} \simeq 6.5 \textrm{k} \Omega.
\end{displaymath}
for all positions $X$ available for the grain\footnote{Note that soft
  matter springs always have some finite thickness even when
  compressed. Coating layers on the leads could also restrict the
  space available for the grain.}. If the criteria for the Coulomb
blockade regime are met, we can consider electrons on the grain to be
fully localized and express the charge on the grain as
\begin{displaymath}
  Q(t) = en(t),
\end{displaymath}
where $n(t)$ takes on only integer values. ($e$ is the electron
charge.) Let $(n,Q_\alpha)$ be the state of the system with $n$ extra
charges on the grain and the charge $Q_\alpha$ on the \mbox{lead
  $\alpha$} ($\alpha={\mathrm L},{\mathrm R})$. It then follows from
the ``orthodox'' Coulomb blockade
theory\cite{a:73_shekhter,a:75_kulik} that the tunneling probabilities
for the tunneling event $(n,Q_{L,R}) \to (n \pm 1, Q_{L,R}\mp e)$ to
occur during the time $\Delta t$ are
\begin{equation}
  \label{eq:tunn_rates}
  \begin{array}{l}
    P_{L,R}^{\pm}(n,X,V,\Delta t) = \nonumber \\ 
    \qquad \Delta t \frac{\Delta G_{L,R}^{\pm}(n,V,X)}{e^2R_{L,R}(X)}
    \left[ 1 - \exp\left(-\frac{\Delta G_{L,R}^{\pm}(n,V,X)}{k_BT}\right)
    \right]^{-1},
  \end{array}
\end{equation}
where $\Delta G_{L,R}^{\pm}(n,V,X)$ is the decrease of free energy in
the system as an electron tunnels to the right ($+$) or to the left
($-$) through the left ($L$) or right ($R$) tunnel junction, $k_B$ is
the Boltzmann constant, and $R_{L,(R)}$ is the resistance of the left
(right) tunnel junction. This resistance depends exponentially on the
displacement of the grain from the center of the system and can be
written
\begin{displaymath}
  R_L(X) = R_R(-X) = R_0 \exp \left( X/\lambda \right),
\end{displaymath}
where $R_0$ is a constant prefactor and $\lambda$ is referred to as
the tunneling length. The tunneling length depends on the materials
used in the system and for our system we estimate $\lambda$ to be of
the order of one Angstrom.

\section{Current-voltage characteristics}
To make the treatment of the model easier we rewrite the equations in
a dimensionless form. If we introduce the dimensionless time
$\tau=t/t_0$ where $t_0=\gamma_d L^2/eV_0$ is a timescale on which the
grain crosses the distance $L$ between the leads due to the
electrostatic forces, the dimensionless length $x=X/L$, the
dimensionless elastic vibration frequency $\omega = \sqrt{kL^2/eV_0}$,
the dimensionless bias voltage $v = V/V_0$ where $V_0=e/4C_0$ is the
Coulomb blockade threshold in the center of the system, and the
dimensionless constant $\alpha = MeV_0 / L^2\gamma_d^2$ which
signifies the ratio between the electrostatic force and a typical
dissipative damping force in the system, we can rewrite the equation
of motion for the grain as:
\begin{equation}
  \alpha\ddot{x} + \dot{x} + \omega^2x = n v + 4 n^2 x
\end{equation}
We will focus on the case when the dissipative force dominates the
electrostatic force while the latter dominates the elastic forces,
$\omega^2 \ll \alpha \ll 1$. The free energy terms $\Delta
G_{L,R}^\pm$ to be used in \mbox{Eq. (\ref{eq:tunn_rates})} are
\begin{eqnarray}
  \label{eq:dg_1}
  \Delta G_L^\pm(n,v,x) = \frac{eV_0}{2} \left(1-4x^2\right)
  \left( -1 \mp 2n \pm \frac{v}{1-2x} \right) \\
  \label{eq:dg_2}
  \Delta G_R^\pm(n,v,x) = \frac{eV_0}{2} \left(1-4x^2\right)
  \left( -1 \pm 2n \pm \frac{v}{1+2x} \right).
\end{eqnarray}
Note that the position dependence of $\Delta G_{L,R}^\pm$ given by
\mbox{Eqs. (\ref{eq:dg_1})} and \mbox{(\ref{eq:dg_2})} results in a
position dependent Coulomb blockade threshold voltage. This means that
whether tunneling in a junction is blocked or not at a certain voltage
depends on where the grain is located at the moment.

\subsection{Numerical approach}
In a numerical approach we have peformed Monte Carlo simulations of
the model system described. A 4:th order Runge Kutta method was used
to solve the equation of motion for the grain for small enough time
steps for the charge on the grain to be considered constant during
each step.  After each time step the charge on the grain was updated
by ``rolling dice'' and deciding whether to carry out a tunneling
event using the tunneling probabilities of \mbox{Eq.
  (\ref{eq:tunn_rates})}.  The current was calculated as the average
of the number of transfered electrons over a certain time interval.
For our choice of parameters the average number of electrons
transfered through the system stabilizes over a time period of
approximately $64 t_0$.  In our calculations of the current - voltage
characteristics we have averaged over $6.4 \cdot 10^3 t_0 = 20\,\mu$s
to reduce the numerical noise. The result of the calculation is
plotted in \mbox{Fig. \ref{fig:iv_comp_inset}}.

If we now compare the current through the studied system with that
through a static symmetric double junction as in \mbox{Fig.
  \ref{fig:iv_comp_inset}} (see inset), it becomes very clear that
there are two distinct parts of the current-voltage curve. For
voltages $V$ approximately between $V_0$ and $1.5 V_0$ the current
through the static symmetric double junction is larger than that
through the system under consideration here. Since, as will be shown
in the next section, charge transport in this regime is dominated by
tunneling, we label this regime \emph{the tunneling regime}. For
voltages above approximately $1.5 V_0$, on the other hand, the current
through the present system is the larger one, and since, as will be
shown below, charge transport in this regime is mechanically mediated
by the grain, we label this regime \emph{the shuttle regime}.

The distinction between the two regimes also becomes very clear if we
consider the root-mean-square of the displacement of the grain from
the center of the system,~$\overline{x}$, as a function of the bias
voltage. This is plotted in \mbox{Fig. \ref{fig:disp_1}}. It is clear
that the average displacement is much larger for the tunneling regime
than for the shuttle regime. The average displacement is also
increasing with the bias voltage for the tunneling regime, whereas,
for the shuttle regime it is a slowly varying function.

\section{Discussion}
\subsection{Tunneling regime}
\label{sec:tunneling_regime}
From \mbox{Fig. \ref{fig:iv_comp_inset}} (inset) we see that for bias
voltages just above the Coulomb blockade threshold $V_0$ (for the
grain in the center position) the current is smaller than it is
through a static symmetric double junction. To understand this we
should consider the $x$-dependence of the tunneling rates. Assume that
the grain starts out sitting uncharged in the center of the system and
that the bias voltage is just above \mbox{$V_0$}. At this point two
things are possible. One unit of charge can either tunnel onto the
grain or off the grain. Since the system is symmetric we consider only
the first of these cases. The criterion for tunneling from the left
lead to the grain is that the free energy is lowered after a tunneling
event, $\Delta G_L^+ > 0$. Using \mbox{Eq.  (\ref{eq:dg_1})} we find
the corresponding inequality
\begin{equation}
  \label{eq:threshold_1}
  x > \frac{1}{2} - \frac{v}{2}.
\end{equation}
Note that $x$ is the normalized coordinate so that \mbox{$-1/2 < x <
  1/2$}.  We see that the Coulomb blockade threshold when the grain is
at the center of the system is ($x=0$) is indeed $V_0$ ($v=1$). For
lower voltages tunneling onto the grain from the \emph{left} lead is
still possible as long as the grain is to the \emph{right} of the
center position. However, this process is exponentially suppressed due
to the increase of resistance with tunneling distance. If one
considers tunneling from a neutral grain to the right lead the same
picture (with $x \to -x$) emerges. When the bias voltage is increased
above $V_0$, tunneling onto the grain becomes possible if it is to the
left of the center. We see here that if the bias voltage is not much
higher than the Coulomb blockade threshold $V_0$, the \emph{open
  region}, where both tunneling onto a neutral grain from the left
lead or off a neutral grain to the right lead is allowed at the same
time, is much smaller than the distance between the leads. The concept
of the open region is illustrated with two examples in \mbox{Fig.
  \ref{fig:open_reg_1}}.

For a grain that has the charge $n=0$ and is located inside the open
region, both the processes $n \to +1$ and $n \to -1$ are allowed at
the same time. If the grain is located outside the open region it can
only be charged from the far lead.

Let us now consider the case when the bias voltage is not much higher
than the Coulomb blockade threshold $V_0$ that applies if the grain is
in the center position. In this case the open region is much smaller
than the distance between the leads. If a unit charge tunnels onto the
grain from the left, the grain becomes positively charged and is thus
affected by a force towards the negative (right) lead.  It will start
to accelerate towards that lead, but if the mass of the grain is very
small and the dissipation large, the grain will reach a maximum
velocity very quickly. As the grain comes close to the negative lead,
the decharging process through the right junction becomes very
probable. If the relaxation of the charge on the grain to the negative
lead takes place outside the open region, the grain cannot be
recharged by a negative unit charge from the negative lead. If
dissipation is strong the grain will stop very quickly and the very
small elastic restoring force will start to move the grain very slowly
towards the center of the system.  At this time the grain is only in
tunneling contact with the far lead and it will continue to move
slowly towards the center, either until it reaches the open region and
can be charged from either lead or until a tunneling event from the
positive lead on the far side of the system occurs again. If the last
of these two processes occurs, the charge on the grain becomes
positive and the grain is accelerated towards the negative lead again,
repeating the above described process. The resulting motion is thus an
oscillation around an average position, which is located between the
open region border and the lead. One cycle of such an oscillation is
schematically illustrated in \mbox{Fig. \ref{fig:open_reg_expl_2}}.

Tunneling from the far lead to the grain as the grain moves under the
influence of the weak restoring force is possible but very unlikely,
as can be seen from \mbox{Eq. (\ref{eq:threshold_1})}. This is due to
the exponential dependence of the tunneling resistance on the
separation between grain and lead. If the grain moves very slowly,
however, there may be enough time for the grain to be charged from the
far lead before it reaches the open region. As the bias voltage is
increased, the size of the open region increases, thereby affecting
the probability that the grain will reach the open region before
getting charged from the far lead. This leads to a transition to the
shuttle regime discussed in the next section.

We can thus conclude that the current through the system is smaller
than that through a static symmetric double junction because the
charge transfer mechanism is limited by tunneling through the more
resistive tunnel barrier, just as is the case for a static asymmetric
double junction. That this is actually the case is also confirmed by
studying plots of the grain position as a function of time, obtained
from Monte Carlo simulations of the system. Such a plot for $V=1.1
V_0$ is shown in \mbox{Fig. \ref{fig:xt_1}}. For clarity the picture
is embedded in a model system with the positions of the leads marked
on the \mbox{$x$-axis}. The plotted line traces out the position of
the grain as a function of time. The sharpness in the curve depends on
the two very different time scales in the system. The time scale for
grain motion due to the electrostatic force when the grain is charged
is much smaller than the time scale of grain motion caused by the weak
elastic force when the grain is uncharged. From the plot we can
conclude that, on average, the charge transfer through the system
looks like that through a static asymmetric double junction.

If we consider \mbox{Eq. (\ref{eq:threshold_1})} and its counterpart,
\begin{displaymath}
  \Delta G_R^+(n=0,x) > 0,
\end{displaymath}
we see that for $V=1.1V_0$, the borders of the open region are located
at {$x=\pm 0.05$}.  When we compare this value to the average
displacement of the grain at this voltage, it becomes clear that the
average displacement is $3$ - $4$ times bigger. The grain thus
performs an oscillatory motion around an average displacement, and
these oscillations are possible because the average displacement is
located quite far away from the border of the open region.

We can now compare the current for this regime, (see {\mbox{Fig.
    \ref{fig:iv_comp_inset}}}) with the current through a static
asymmetric double junction. The current through the latter type of
double junction can be approximated by saying that the charge transfer
to the far lead limits the current. Since the inverse of the tunneling
rate is the average time between tunneling events, we can write the
current $I_{a.d.j.}$ through the asymmetric double junction as
\begin{equation}
  I_{a.d.j.} = e\Gamma_{far-lead},
\end{equation}
where $\Gamma_{far-lead}$ is the rate for tunneling events between the
grain and the far lead. Using \mbox{Eq.
  (\ref{eq:tunn_rates})\cite{comment:1}} under, for instance, the
assumptions $x>0$ and $T=0$, the current from the far lead would be
\begin{equation}
  \label{eq:adj_curr_1}
  I_{a.d.j.} = \frac{e}{8R_0C_0}
  \frac{\left( 1 - 4x^2 \right)\left( -1 + \frac{v}{1-2x} \right)}
  {\exp \left( \frac{L}{\lambda}x \right)}.
\end{equation}
If we use the average displacement from \mbox{Fig. \ref{fig:disp_1}},
the current, as calculated by \mbox{Eq. (\ref{eq:adj_curr_1})} and in
the voltage interval $1<V/V_0<1.25$, turns out to be of the order
\mbox{$20$ \%} lower than the actual current through the our system.
This is understandable since the small grain oscillations around the
average displacement decrease the effective tunneling resistances seen
by the charges transfered through the system.

\subsection{Shuttle regime}
\label{sec:shuttle_regime}
The statements made in the previous section mean that we can expect
the current through our device to be very small on the scale of the
current through a symmetric static double junction. On this scale, we
can also expect that the current only increases slowly as the bias
voltage is raised to slightly above the Coulomb blockade threshold in
the center of the system. The current will continue to increase very
slowly with the voltage. As the size of the open region increases it
becomes more and more probable that an empty grain will reach the open
region before it is recharged from the far lead. If the grain reaches
the open region, charge transfer from the near lead suddenly becomes
the dominating charge transfer mechanism. If we consider \mbox{Eq.
  (\ref{eq:threshold_1})} we see that as the bias voltage $V$ reaches
$2 V_0$, the open region has extended all the way to the leads. The
grain will thus always move inside the open region. In this case, the
charge transfer cycle looks quite different from the picture in the
previous section. When the grain gets positively charged it will move
towards the negative lead. As the grain gets closer to the lead, the
tunnel resistance decreases exponentially and finally the charge on
the grain will tunnel from the grain to the lead. When the grain loses
its charge it will stop very quickly due to the high dissipation. The
grain now starts to move very slowly towards the center of the system,
but since the timescale of charge exchange with the near lead is much
shorter than that of movement due to the elastic force, another tunnel
event can occur and the grain can get negatively charged. This means
that the grain will be accelerated towards the positive lead, where a
similar procedure will occur. The grain will now continue to move back
and forth in this fashion, shuttling charge across the junction.
\mbox{Figure \ref{fig:shut_reg_1}} shows a schematic illustration of
this charge transfer mechanism.

The exponentially large tunnel resistances limiting the current in the
tunneling regime are now gone, since all tunneling events occur when
the grain is close to the leads. The oscillations of the grain thus
effectively lower the tunnel barriers seen by the transfered charges,
which leads to a large increase in the current.

We can now proceed as in the case of the tunneling regime and consider
a plot of the grain position as a function of time for some bias
voltage in this interval. In \mbox{Fig. \ref{fig:xt_2}} we have made
such a plot for the bias voltage $V=2.0 V_0$. As in \mbox{Fig.
  \ref{fig:xt_1}} the plot is shown together with the model system so
that the positions of the leads are marked on the \mbox{$x$-axis}. The
plotted line traces out the position of the grain as a function of
time. The grain performs a stochastic but still oscillatory motion
back and forth through the system. For this voltage an uncharged grain
is everywhere in tunneling contact with both leads so that the charge
on the grain can change by $2e$ at each approach of a lead. This means
that the grain will always be pushed by the electrostatic force, which
explains the much shorter time scale for grain motion in \mbox{Fig.
  \ref{fig:xt_2}} compared to \mbox{Fig. \ref{fig:xt_1}} (Note also
the factor of $10$ difference in scale on the time axes in the two
plots).

Let us now go back and consider the $IV$-curve in \mbox{Fig.
  \ref{fig:iv_comp_inset}} again. For a bias voltage of approximately
$4.5 V_0$, the $IV$-curve changes slope over a relatively short
voltage interval. The reason for this is the transition to a regime
where two extra charges are allowed on the grain, i.e. four charges
can be transported across the system in each shuttle cycle. To get a
better understanding of this, we should consider the case $\Delta
G_L^+(n=1,x) > 0$, i.e. the condition that the free energy decrease
should be positive when one charge tunnels from the left lead onto an
already charged grain.  Using \mbox{Eq. (\ref{eq:dg_1})}, we get the
condition
\begin{equation}
  \label{eq:threshold_2}
  x > \frac{1}{2}-\frac{v}{6}.
\end{equation}
We thus see that when $V=3 V_0$ ($v=3$) a new open region develops,
where electron tunneling is allowed from the left lead when the grain
charge is $n=1$ and to the right lead when the grain charge is $n=-1$.
If we remind ourselves of what went on in the tunneling regime, we
cannot expect that the current will change much until the size of this
region is of the same size as the amplitude of the grain oscillations.
As can be seen from \mbox{Fig.  \ref{fig:iv_comp_inset}}, nothing new
happens to the $IV$-characteristics when $V=3 V_0$. However,
approximately when $V=4.5 V_0$, there is a transition to the new
regime. From \mbox{Eq.  (\ref{eq:threshold_2})} we get that, at
$v=4.5$, the open region borders for $n=1$ have extended to
approximately $x \in (-0.25,0.25)$.  At this voltage, the grain
oscillations should thus be inside the new region most of the time,
allowing the transfer of four charges in each shuttle cycle.

It is important to note here that the transition in the $IV$-curve is
not sharp. As the open region for $n=1$ grows wider, it will become
more and more probable that, as the grain moves across the system, it
will transport two charges instead of only one charge. As is normally
the case for shuttle transport\cite{a:98_gorelik}, we can consider a
current frequency relationship:
\begin{displaymath}
  I=2{\cal N}ef \,,
\end{displaymath}
where ${\cal N}$ is defined by this equation and represents an average
number of extra electrons transported on the grain, and where $f$ is
the vibrational frequency of the grain. Both ${\cal N}$ and $f$ are
functions of the bias voltage. Note also that ${\cal N}$ is not
usually an integer.

\subsection{Analytical description of the shuttle regime}
\label{sec:an_desc_shuttle_reg}
In this section we present an analytical approach to modeling the
current through the system for bias voltages in the range $2<V/V_0<3$.
In this voltage interval, the grain can only shuttle one charge at a
time in each direction. Since the motion of the grain is strongly
influenced by the random tunneling events, we have to consider the
period time in an averaged sense and write the current as:
\begin{equation}
  I = \frac{2e}{t_0 \overline{T}},
\end{equation}
where $\overline{T}$ is the dimensionless average oscillation period
and $t_0=\gamma_d L^2/eV_0$ is the typical time scale in the system.
We make the assumption that we can divide the average period into the
three parts schematically illustrated in \mbox{Fig. \ref{fig:ill_1}}.
Since the system is symmetric with respect to the center of the system
it is enough to consider half a cycle.

The first part of the average period, $T_1$, is the average time it
takes a grain with one excess charge to move from the center of the
system towards the negative lead to the position where, on average,
the excess charge is relaxed to the negative lead. After the charge
has relaxed to the negative lead, the grain stops very quickly and, on
the average, sits still during the time $T_2$ before one more charge
tunnels to the negative lead. As this happens, it takes the grain the
time $T_3$ to get back to the center of the system, where it repeats a
mirror version of this cycle towards the positive lead. Note also that
the further the grain moves towards the lead, the shorter the time
$T_2$ can be expected to be. As the grain passes the center of the
system towards one lead, there is at each position a certain
probability that the charge on the grain will tunnel to the lead.
Wherever the tunneling event occurs, the two average times $T_2$ and
$T_3$ are determined by the first time $T_1$, which is determined by
the position at which the tunneling event occurred. We can therefore
write the average period time as: 
\begin{equation}
  \overline{T} = 2\int_0^{x_{max}} \tau(x) P(x) dx,
\end{equation}
where $\tau(x)$ is the half-period for a grain that reaches position
$x$ as it travels from the center of the system towards the lead. This
half-period now consists of the sum of three partial times
$\tau_1(x)$, $\tau_2(x)$ and $\tau_3(x)$, where the indexes refer to
the same parts of the half-period as the time indexes illustrated in
\mbox{Fig. \ref{fig:ill_1}}.

To find the probability density $P(x)$, we can consider an ensemble
consisting of $N$ grains. These grains all start out at the center of
the system, have charge $n=1$ and move towards the negative lead. We
can first find the relative number of grains $m(x)/N$ that still has a
charge of $n=1$ at $x$ by noting that:
\begin{equation}
  \frac{d\left( \frac{m(x)}{N} \right)}{dx} = 
  -\frac{m(x)}{N} \frac{\Gamma_R^+(n=1,x)}{\dot{x}}.
\end{equation}
This is an ordinary separable differential equation with the solution:
\begin{equation}
  \frac{m(x)}{N} = \frac{m(0)}{N} \exp \left( -\int_0^x
  \frac{t_0\Gamma_R^+(n=1,x')}{\dot{x}(x')} dx' \right).
\end{equation}
Since all grains in the ensemble have charge $n=1$ at $\tau=0$, we see
that $m(0)/N=1$. We can now find the probability density $P(x)$ as the
relative number of grains in the ensemble that stops at precisely $x$,
i.e. minus the derivative of $m(x)/N$;
\begin{equation}
  \label{eq:prob_1}
  P(x) = \frac{t_0\Gamma_R^+(n=1,x)}{\dot{x}(x)} \exp 
  \left( -\int_0^x \frac{t_0\Gamma_R^+(n=1,x')}{\dot{x}(x')} dx'\right).
\end{equation}
The next step is to find the half-period, $\tau(x)$. Since we are
working in the high dissipation limit, $\alpha \ll 1$, acceleration
times are very short compared to the time scales of movement of the
grain and tunneling.  This means that, we can to a good approximation
find the parts $\tau_1(x)$ and $\tau_3(x)$ by integrating the equation
for the velocity of the grain\cite{comment:2}:
\begin{equation}
  \dot{x} = n v + 4 n^2 x,
\end{equation}
from $\tau=0$ to $\tau=\tau'(x)$, and for $n = \pm 1$. The resulting
traveling times are:
\begin{eqnarray}
  \label{eq:t1}
  \tau_1(x) &=& \frac{1}{4}\ln \left( 1 + \frac{4x}{v} \right) \\
  \label{eq:t3}
  \tau_3(x) &=& -\frac{1}{4}\ln \left( 1 - \frac{4x}{v} \right).
\end{eqnarray}

To find the time $\tau_2$ we first assume that the grain will not move
on the scale of the tunneling length during this time. This means that
the tunneling rates are time independent and that we, if the grain
sits with zero charge at $x$, can expect the average time before a
tunneling event occurs to be:
\begin{equation}
  \tau_2(x) = \frac{1}{t_0 \Gamma_R^+(n=0,x)}.
\end{equation}
At zero temperature we can expect the time to be:
\begin{equation}
  \label{eq:t2}
  \tau_2(x) = \frac{8R_0C_0}{t_0} \frac{\exp \left
  ( -\frac{L}{\lambda}x \right)}
  {\left( 1 - 4x^2 \right)\left( -1 + \frac{v}{1+2x} \right)}.
\end{equation}
We have thus arrived at the following expression for the current
through the system in the bias voltage interval $2<V/V_0<3$:
\begin{equation}
  \label{eq:curr_1}
  I = \frac{e}{t_0\int_0^{x_{max}}
    \left( \tau_1(x)+\tau_2(x)+\tau_3(x)\right) P(x) dx},
\end{equation}
where $\tau_1(x)$, $\tau_2(x)$ and $\tau_3(x)$ are given by the
equations (\ref{eq:t1}), (\ref{eq:t2}) and (\ref{eq:t3}) and $P(x)$ is
given by \mbox{Eq. (\ref{eq:prob_1})}.

We have, with the same parameters as used in our earlier Monte Carlo
simulations, numerically calculated the current given by \mbox{Eq.
  (\ref{eq:curr_1})}. The results are shown in \mbox{Fig.
  \ref{fig:comp_2}}. The solid line corresponds to the Monte Carlo
simulations of the system and the circles correspond to the values
obtained from \mbox{Eq. (\ref{eq:curr_1})}. The agreement between the
numerical studies and the analytical approach is very good, which is a
strong indication that the charge shuttle mechanism description of the
charge transfer is applicable also in highly dissipative systems.

It is also of interest to know the threshold voltage, $V_t$, and the
width, $\Delta V$, of the transition from the tunneling regime to the
shuttle regime. In order to estimate these we can consider small
oscillations, $\Delta x$, of the grain around some average position
$x_0$. Without loss of generality we can assume that $x_0>0$, i.e. the
grain oscillates on the right hand side of the system. If we assume
that the oscillation amplitudes are not very big we can estimate the
velocity of the grain to be
\begin{equation}
  v_{IN} \approx -\omega^2 x_0,
\end{equation}
for grains moving towards the center of the system due to the elastic
force. On average it moves during the time
\begin{equation}
  \tau_{IN} \approx \frac{1}{t_0 \Gamma_L^+(n=0,v,x_0)},
\end{equation}
before it is charged from the far lead. When the grain is moving
towards the lead due to the electrostatic force acting on the extra
charge on the grain, it approximately moves with the velocity
\begin{equation}
  v_{OUT} \approx (v+4x_0).
\end{equation}
The average time it will move before the extra charge tunnels to the
right lead is
\begin{equation}
  \tau_{OUT} \approx \frac{1}{t_0 \Gamma_R^+(n=1,v,x_0)}.
\end{equation}
For the position $x_0$ to be stable the average the distance the grain
moves in each direction has to be equal to each other. We thus get the
relation:
\begin{equation}
  \label{eq:dx_relation_1}
  \frac{\omega^2 x_0}{f_L}e^{\frac{L}{\lambda}x_0} =
  \frac{(v+4x_0)}{f_R}e^{-\frac{L}{\lambda}x_0},
\end{equation}
where $f_L = -1/2 + C_RV/e$ and $f_R=1/2 + C_LV/e$ are functions of
the right and left capacitances and the bias voltage and that are of
order unity as long as the grain is not close to the open region
border. Rearranging the factors in \mbox{Eq. (\ref{eq:dx_relation_1})}
and taking the logarithm of both sides we get
\begin{equation}
  \label{eq:dx_relation_2}
  2\frac{L}{\lambda}x_0 = \ln\frac{1}{\omega^2} +
  \ln\frac{v+4x_0}{x_0} + \ln\frac{f_R}{f_L}.
\end{equation}
Under the conditions that we are not close to the open region border
and that the elastic force is very weak we can neglect the last two
terms on the right hand side of \mbox{Eq. (\ref{eq:dx_relation_2})}.
In this case we get the average position for the grain as
\begin{equation}
  x_0 \approx - \frac{\lambda}{2L}\ln\omega^2.
\end{equation}
If $\Delta x \ll x_0$, one can, by comparing the average position,
$x_0$, for the grain with the open region border, $(v-1)/2$, estimate
the threshold voltage,
\begin{equation}
  V_t = V_0\left(1-\frac{\lambda}{L} \ln \frac{kL^2}{eV_0}\right),
\end{equation}
which corresponds to the transition from the tunneling regime to the
shuttle regime.

We can now use the expression for $x_0$ to estimate the width of the
oscillations as
\begin{equation}
  \Delta x \sim 2
  \frac{\lambda}{L}
  \frac{R_0C_0}{t_0}
  \sqrt{\omega^2}
  \ln \frac{1}{\omega^2}.
\end{equation}
From \mbox{Eq. (\ref{eq:threshold_1})} we know that the open region
expands linearly with the bias voltage. When the oscillations are
completely outside the open region we can expect the system to be in
the tunneling regime. When the open region has expanded to include the
oscillations, the system should be in the shuttle regime. The open
region border expands $\Delta x$ if the voltage is increased with
$\Delta V / V_0 = 2 \Delta x$ and we thus get the relative transition
width as:
\begin{eqnarray}
  \frac{\Delta V}{V_t-V_0} = 
  \frac{\Delta x}{x_0} &\sim&
  4 \frac{R_0C_0}{t_0} \sqrt{\omega^2} \nonumber \\
  \label{eq:transition_width}
  &=& 4\frac{\omega_{sh}}{\omega_R}\eta^{-\frac{1}{2}},
\end{eqnarray}
where $\omega_{sh}=t_0^{-1} = eV_0/\gamma_dL^2$ is a typical grain
oscillation frequency, $\omega_R=1/R_0C_0$ is a characteristic
tunneling frequency, and $\eta = \omega^2 = kL^2/eV_0$ represents the
strength of the electromechanical coupling. From \mbox{Eq.
  (\ref{eq:transition_width})} one can see that there are two cases
when there is a very sharp transition between the two regimes. The
first case is when the electromechanical coupling becomes very strong.
The second case is when the shuttle frequency is low compared to the
rate of tunneling. In our system these conditions are realized by the
assumed weak elastic forces and the high rate of dissipation
associated with the moving grain.

\section{Conclusions}
The main conclusion resulting from our analysis is that an
electromechanical coupling in dissipative nanometer sized Coulomb
blockade structures cannot be viewed simply as an additional channel
for absorbing the power associated with the current injected into the
system. Instead a new mechanism of mechanically assisted charge
transfer occurs, which increases the current exponentially and which
to some extent is related to the shuttling of electrical charges,
predicted for weakly dissipative electromechanical
structures\cite{a:98_gorelik,a:98_isacsson}. We have shown that the
electromechanical coupling results in a highly nonlinear IV-curve with
two distinct regimes of charge transport. More features of the charge
transfer might be available by studying the noise properties of the
system. Since the noise is sensitive to the dynamical properties of
the system, noise measurements can give additional information about
the interplay between elasticity and dissipation in real
nanoelectromechanical structures.

\acknowledgments 
Financial support from the Swedish TFR/VR (T.N.), SSF through the QDNS
program (L.G.), and NFR/VR (R.S.) is gratefully acknowledged.

\bibliographystyle{plain}

\begin{thebibliography}{99}
\bibitem{comment:2}This equation is valid under the assumption that
  the acceleration time is much shorter than the timescale of the
  motion of the grain, which is the case due to the high dissipation
  in contrast to the weak elastic force.
\end{thebibliography}

\newpage
\onecolumn
\begin{center}
  
  \begin{figure}[h]
    \begin{center}
      \psfrag{r1}{\small$\mathbf{R_L}$}
      \psfrag{r2}{\small$\mathbf{R_R}$}
      \psfrag{+V/2}{\hspace*{1mm}$\mathbf{+\frac{V}{2}}$}
      \psfrag{-V/2}{$\mathbf{-\frac{V}{2}}$}
      \psfrag{k}[]{\small$\mathbf{k}$}
      \psfrag{m}[]{\hspace*{2mm}\small$\mathbf{M}$}
      \psfrag{L}[]{\small$\mathbf{L}$}
      \includegraphics[scale=0.4]{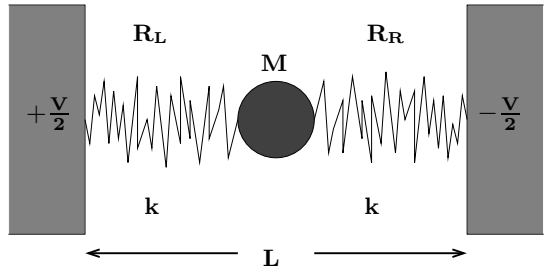}
      \vspace{5 mm}
      \caption{Schematic picture of the model system, which consists of a
        metallic grain of mass $M$ coupled by weak elastic links to
        two electrodes separated by a distance $L$. The elastic links
        act as springs with spring constant $k$. The tunneling
        resistances of the right and left junctions are $R_R$ and
        $R_L$. A bias voltage $V$ is applied across the system.}
      \label{fig:system}
    \end{center}
  \end{figure}

  \newpage
  \begin{figure}[htbp]
    \begin{center}
      \psfrag{I (nA)}[b]{\hspace*{2mm}\raisebox{2mm}{\small$\mathbf{I (nA)}$}}
      \psfrag{V}[t]{\hspace*{2mm}\small$\mathbf{V/V_0}$}
      \includegraphics[scale=0.4]{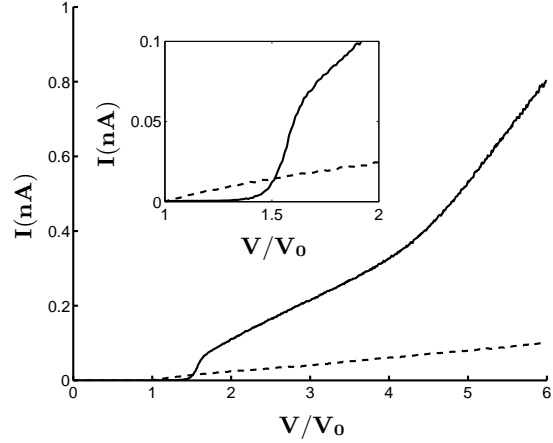}
      \vspace{5 mm}
      \caption{The solid line shows the current - voltage
        characteristics obtained by a Monte Carlo simulation of the
        charge transport through the system sketched in
        \mbox{Fig.\ref{fig:system}}.  The calculated current, which
        was averaged over $20 \mu$s, is plotted as a function of the
        bias voltage $V$ scaled by the Coulomb blockade theshold
        voltage $V_0$, that applies if the movable grain is equally
        far from both electrodes.  The dashed line displays the
        current through a static symmetric double junction for the
        same parameters. The parameters used in the simulation are:
        $\alpha = 6.4*10^{-4}$ and $\omega^2=4.27*10^{-3}$. It is
        clear that for voltages between approximately $V_0$ and $1.5
        V_0$ (see the inset which shows a magnification of the voltage
        interval $1<V/V_0<2$) the current through the model system is
        smaller than the current through the static symmetric double
        junction, whereas, for higher voltages it is the other way
        around.}
      \label{fig:iv_comp_inset}
    \end{center}
  \end{figure}

  \newpage
  \begin{figure}[htb]
    \begin{center}
      \psfrag{disp}[]{\raisebox{7mm}{\textbf{RMS Displ. ($\overline{X}/L$)}}}
      \psfrag{V}[]{\hspace*{2mm}\raisebox{-5mm}{\small$\mathbf{V/V_0}$}}
      \includegraphics[scale=0.4]{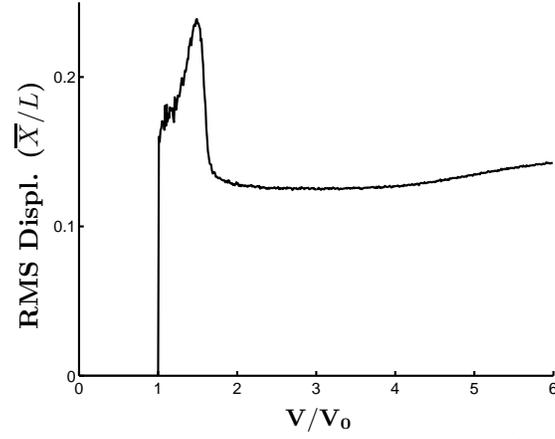}
      \vspace*{3mm}
      \caption{The root-mean-square displacement of the grain from the
        symmetric position between the leads as a function of the bias
        voltage scaled by $V_0$, the Coulomb blockade threshold
        voltage in the center of the system. The parameters used in
        the simulation are: $\alpha = 6.4*10^{-4}$ and
        $\omega^2=4.27*10^{-3}$.  The distinction between the two
        different regimes of charge transfer is very clear. In the
        tunneling regime, the average displacement increases with the
        voltage and is larger than in the shuttle regime. In the
        shuttle regime, the average displacement is a slowly varying
        function of the voltage.}
      \label{fig:disp_1}
    \end{center}
  \end{figure}

  \newpage
  \begin{figure}[htb]
    \begin{center}
      \raisebox{25mm}{(a)}
      \begin{minipage}[t]{0.4\textwidth}
        \psfrag{n+}{\small$\mathbf{n \to +1}$}
        \psfrag{n-}{\small$\mathbf{n \to -1}$}
        \psfrag{+V/2}{\small\hspace*{2mm}$\mathbf{+\frac{V}{2}}$}
        \psfrag{-V/2}{\small\hspace*{1mm}$\mathbf{-\frac{V}{2}}$}
        \psfrag{n}[]{\raisebox{2.5mm}{\small\hspace*{3.5mm}$\mathbf{n=0}$}}
        \includegraphics[scale=0.45]{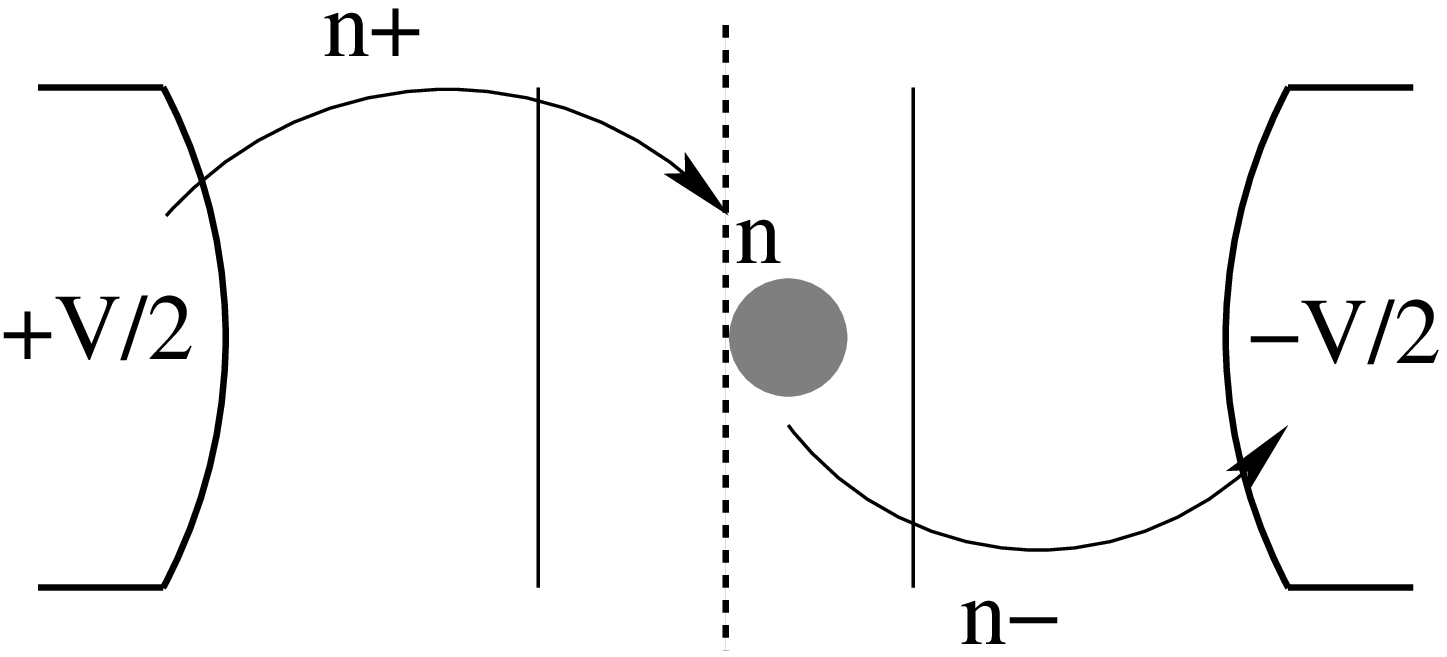}
      \end{minipage}
      \vspace*{8mm}\\
      \raisebox{25mm}{(b)}
      \begin{minipage}[b]{0.4\textwidth}
        \psfrag{n+}{\small$\mathbf{n \to +1}$}
        \psfrag{n2+}{\small$\mathbf{n \to -1}$}
        \psfrag{+V/2}{\small\hspace*{2mm}$\mathbf{+\frac{V}{2}}$}
        \psfrag{-V/2}{\small\hspace*{1mm}$\mathbf{-\frac{V}{2}}$}
        \psfrag{n}[]{\raisebox{2mm}{\small\hspace*{2.5mm}$\mathbf{n=0}$}}
        \psfrag{n2}[]{\raisebox{2mm}{\small\hspace*{2.5mm}$\mathbf{n=0}$}}
        \includegraphics[scale=0.45]{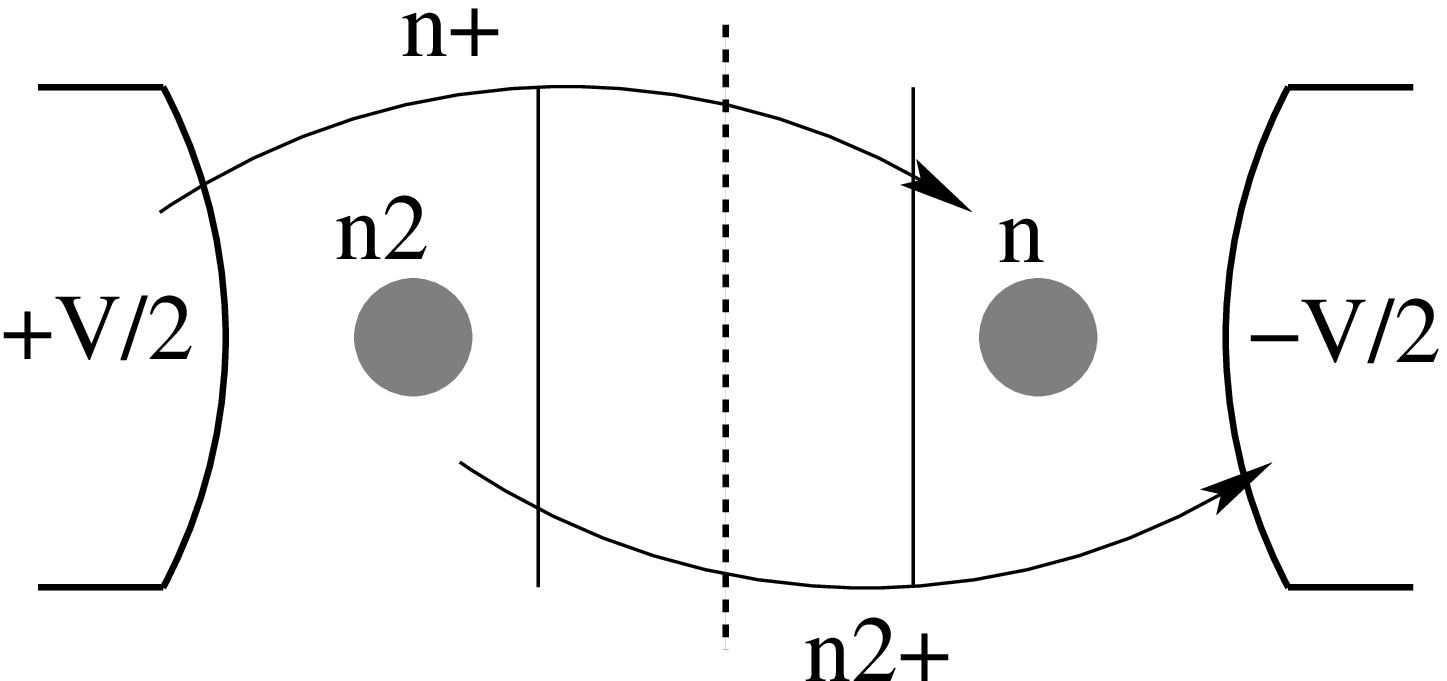}
      \end{minipage}
      \vspace*{3mm}
      \caption{Illustration of the concept of the open region which, in
        the pictures above, correspond to the space between the
        vertical solid lines. (a) If the grain is uncharged and
        located inside the open region, it is in tunneling contact
        with both leads at the same time. (b) When the grain is
        situated outside the open region, energy considerations show
        that tunneling to the near lead is blocked. Tunneling from the
        far lead is still possible, however, this process is strongly
        suppressed due to the exponential dependence of the tunneling
        resistance on the grain-lead separation.}
      \label{fig:open_reg_1}
    \end{center}
  \end{figure}

  \newpage
  \begin{figure}[tb]
    \begin{center}
      \psfrag{n+}{\small\hspace*{-5mm}$\mathbf{n \to +1}$}
      \psfrag{+V/2}{\small\hspace*{1mm}$\mathbf{+\frac{V}{2}}$}
      \psfrag{-V/2}{\small\hspace*{1mm}$\mathbf{-\frac{V}{2}}$}
      \psfrag{n}[]{\raisebox{2mm}{\small\hspace*{5mm}$\mathbf{n=0}$}}
      \includegraphics[scale=0.45]{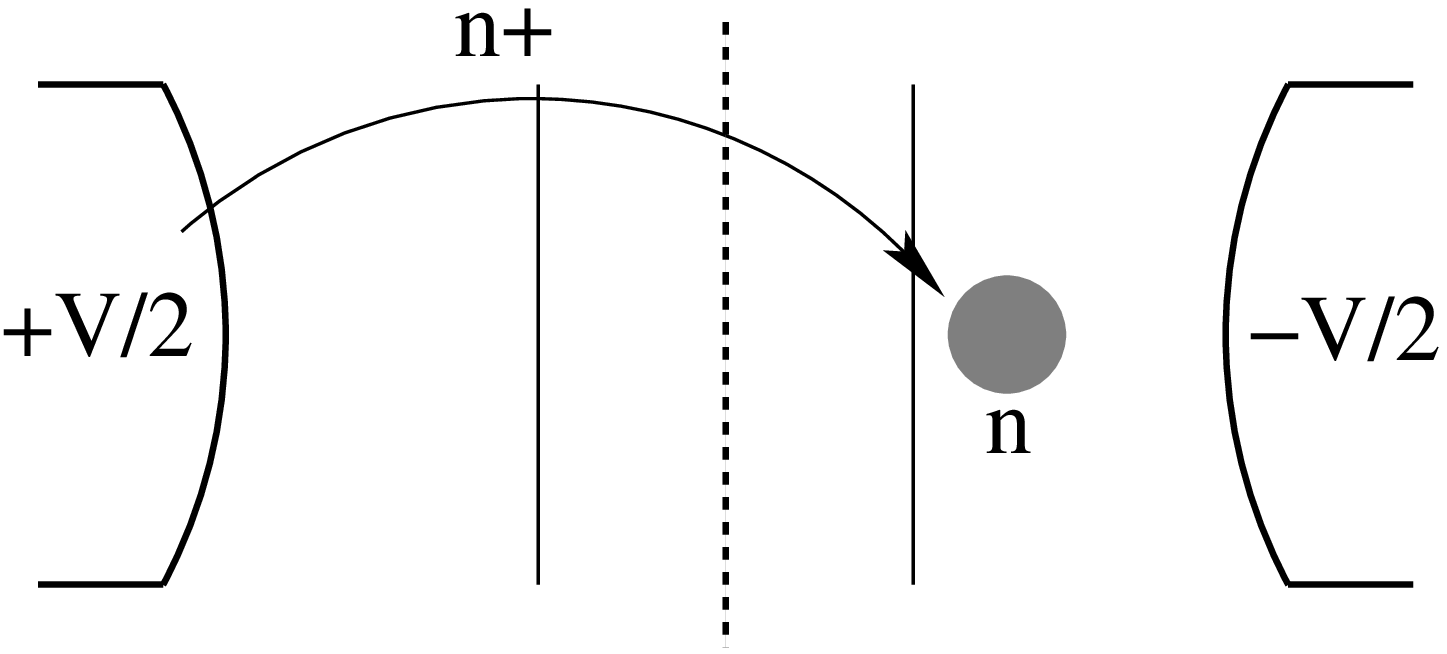}\\
      \begin{minipage}{0.45\textwidth}
        1: The grain is charged from the far lead.
      \end{minipage}
      \vspace*{5mm}\\
      \psfrag{n+}[]{\raisebox{-8mm}{\small\hspace*{-8mm}$\mathbf{n \to 0}$}}
      \psfrag{+V/2}{\small\hspace*{1mm}$\mathbf{+\frac{V}{2}}$}
      \psfrag{-V/2}{\small\hspace*{1mm}$\mathbf{-\frac{V}{2}}$}
      \psfrag{n}[]{\raisebox{2mm}{\small\hspace*{4mm}$\mathbf{n=1}$}}
      \includegraphics[scale=0.45]{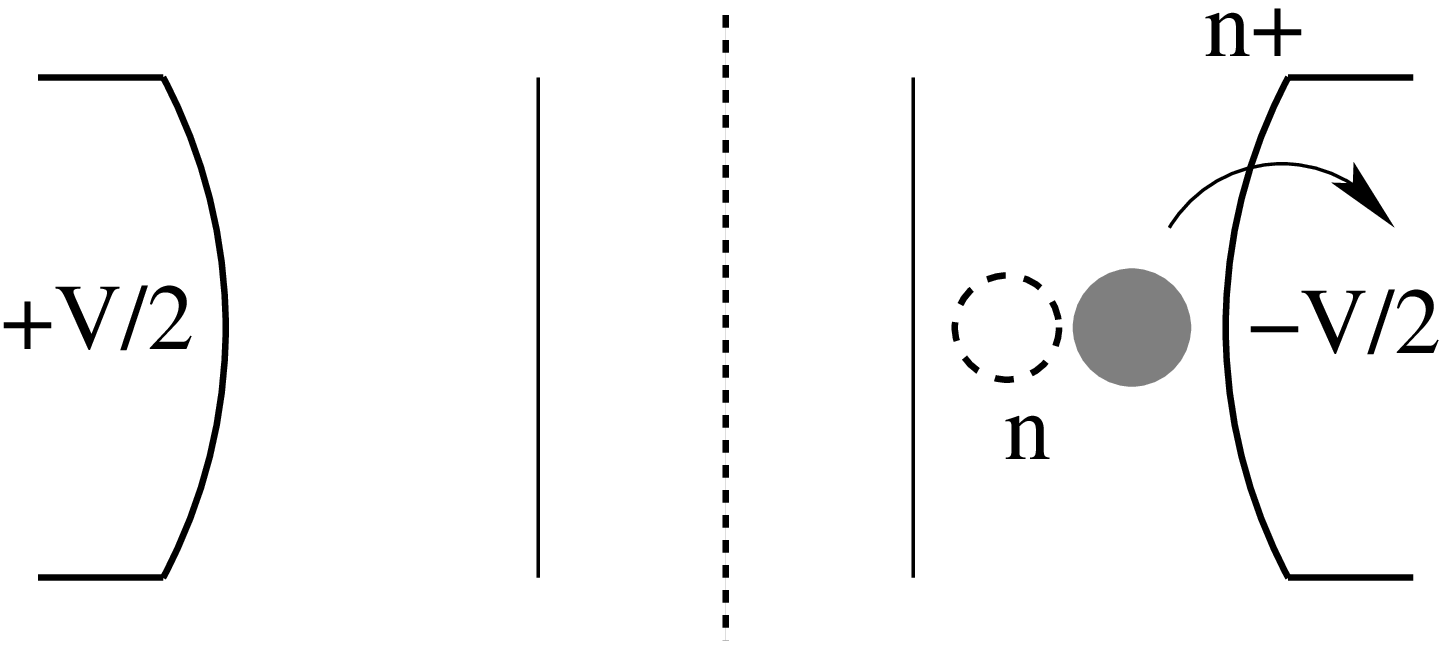}\\
      \begin{minipage}{0.45\textwidth}
        2: Being charged, the grain is pushed towards the near lead
        until a tunnel event to that lead occurs.
      \end{minipage}
      \vspace*{5mm}\\
      \psfrag{n+}{\small\hspace*{-5mm}$\mathbf{n \to +1}$}
      \psfrag{+V/2}{\small\hspace*{1mm}$\mathbf{+\frac{V}{2}}$}
      \psfrag{-V/2}{\small\hspace*{1mm}$\mathbf{-\frac{V}{2}}$}
      \psfrag{n}[]{\raisebox{2mm}{\small\hspace*{4mm}$\mathbf{n=0}$}}
      \includegraphics[scale=0.45]{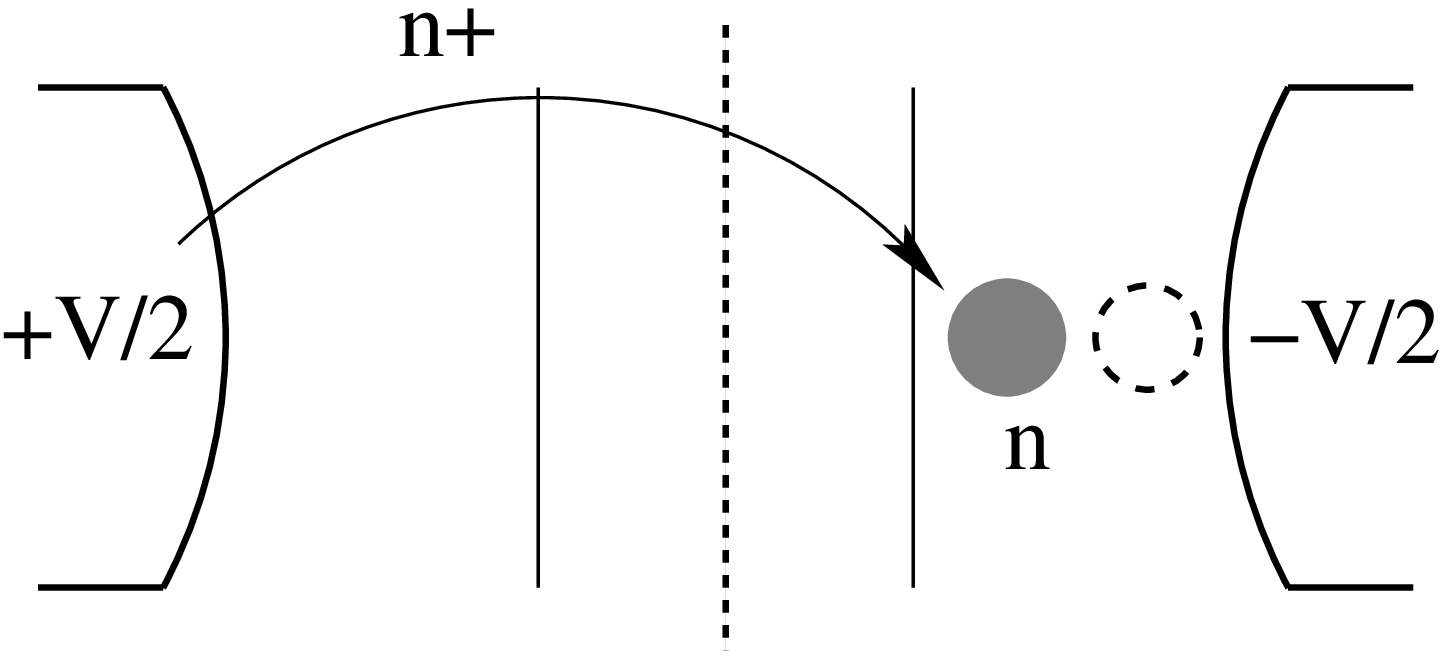}\\
      \begin{minipage}{0.45\textwidth}
        3: The grain is slowly pulled back by the weak elastic force
        until charging from the far lead occurs again.
      \end{minipage}
    \end{center}
    \caption{Schematic illustration of the charge transfer mechanism in
      the tunneling regime. The grain performs small oscillations
      around an average position, located between the the open region
      border and the lead. In the figures above, the open region is
      bounded by the vertical solid lines in the center of each
      junction.}
    \label{fig:open_reg_expl_2}
  \end{figure}

  \newpage
  \begin{figure}[htb]
    \begin{center}
      \psfrag{pos}{\hspace*{-3mm}\raisebox{-4mm}{\textbf{POSITIVE LEAD}}}
      \psfrag{neg}{\hspace*{-4mm}\textbf{NEGATIVE LEAD}}
      \psfrag{0}[]{}
      \psfrag{1}[]{\hspace*{2mm}\raisebox{-1mm}{\small{$\mathbf{1.0}$}}}
      \psfrag{2}[]{\hspace*{2mm}\raisebox{-1mm}{\small{$\mathbf{2.0}$}}}
      \psfrag{3}[]{\hspace*{2mm}\raisebox{-1mm}{\small{$\mathbf{3.0}$}}}
      \psfrag{4}[]{\hspace*{2mm}\raisebox{-1mm}{\small{$\mathbf{4.0}$}}}
      \psfrag{5}[]{\hspace*{2mm}\raisebox{-1mm}{\small{$\mathbf{5.0}$}}}
      \psfrag{6}{}
      \psfrag{t}{\small$\mathbf{t}$ $\mathbf{\mu}$s}
      \psfrag{x}{\hspace{-5mm}$\mathbf{X(t)/L}$}
      \psfrag{-}{\hspace{-2mm}$-\frac{1}{2}$}
      \psfrag{+}{\hspace{-2mm}$+\frac{1}{2}$}
      \includegraphics[scale=0.3]{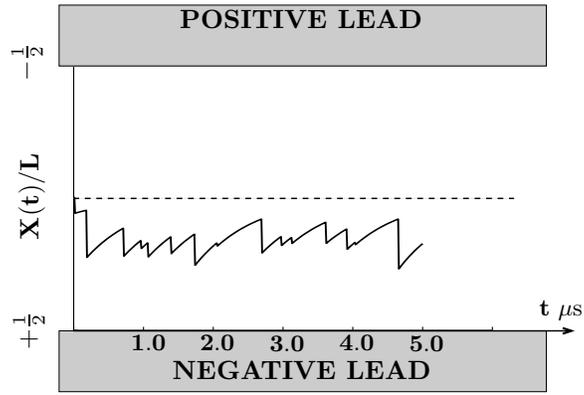}
      \vspace*{5mm}
      \caption{Plot of the position of the grain as a function of time
        for the bias voltage $V=1.1 V_0$ and the parameters $\alpha =
        6.4*10^{-4}$ and $\omega^2=4.27*10^{-3}$. For this voltage the
        behavior of the system is on average very much like a static
        symmetric double junction. The jaggedness of the curve comes
        from the very different velocities of charged and uncharged
        grains.}
      \label{fig:xt_1}
    \end{center}
  \end{figure}
  
  \newpage
  \begin{figure}[htb]
    \begin{center}
      \psfrag{n+}{\small\hspace*{-1mm}$\mathbf{n \to +1}$}
      \psfrag{+V/2}{\small\hspace*{1mm}$\mathbf{+\frac{V}{2}}$}
      \psfrag{-V/2}{\small\hspace*{1mm}$\mathbf{-\frac{V}{2}}$}
      \psfrag{n}[]{\raisebox{1mm}{\small\hspace*{5mm}$\mathbf{2e}$}}
      \includegraphics[scale=0.45]{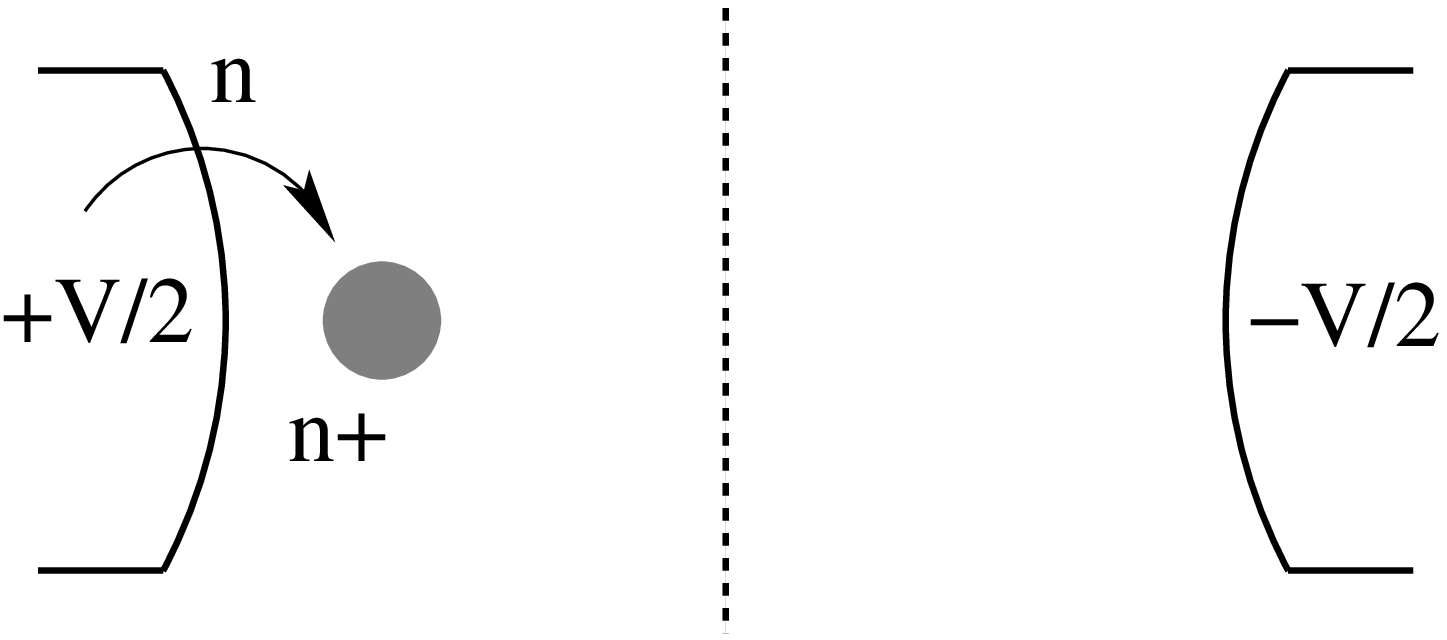}\\
      \begin{minipage}{0.45\textwidth}
        1: The grain is charged from the near lead.
      \end{minipage}
      \vspace*{5mm}\\
      \psfrag{n+}{\small\hspace*{-1mm}$\mathbf{n \to -1}$}
      \psfrag{+V/2}{\small\hspace*{1mm}$\mathbf{+\frac{V}{2}}$}
      \psfrag{-V/2}{\small\hspace*{1mm}$\mathbf{-\frac{V}{2}}$}
      \psfrag{n}[]{\raisebox{2mm}{\small\hspace*{8mm}$\mathbf{2e}$}}
      \includegraphics[scale=0.45]{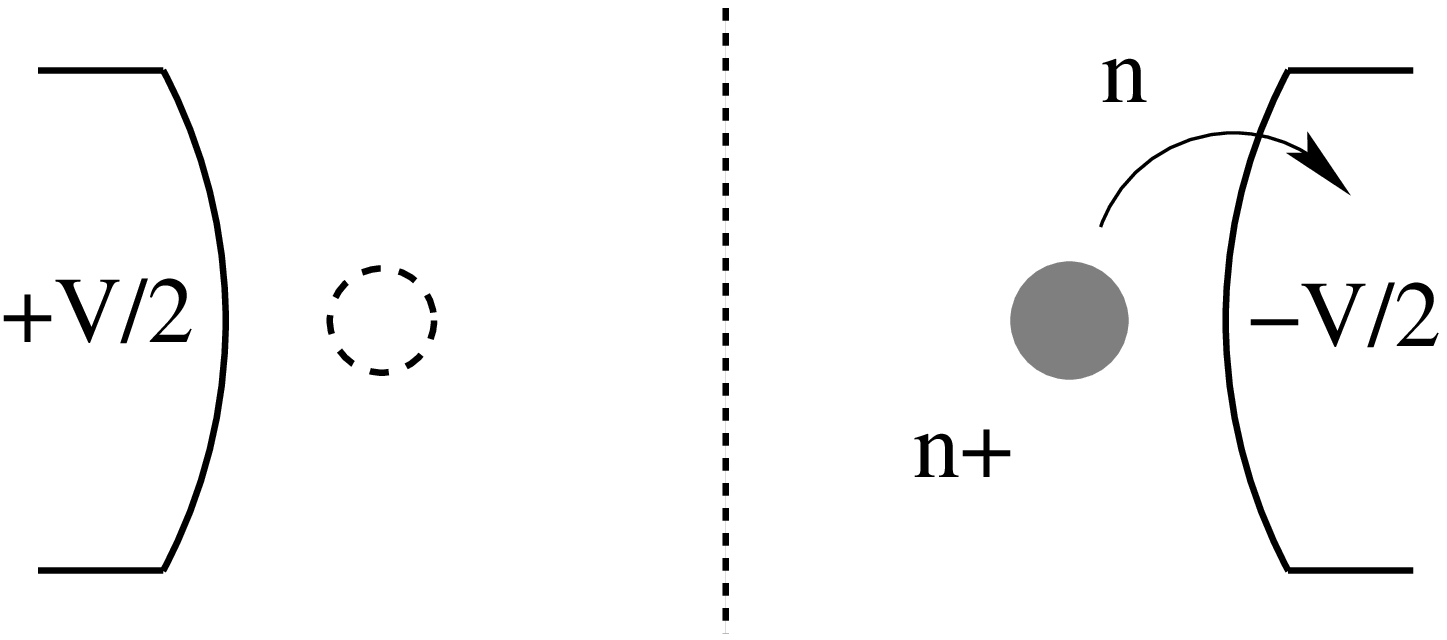}\\
      \begin{minipage}{0.4\textwidth}
        2: Being positively charged, the grain is pushed towards the
        other lead, where two charges tunnel off the grain.
      \end{minipage}
      \vspace*{5mm}\\
      \psfrag{n+}{\small\hspace*{-1mm}$\mathbf{n \to +1}$}
      \psfrag{+V/2}{\small\hspace*{1mm}$\mathbf{+\frac{V}{2}}$}
      \psfrag{-V/2}{\small\hspace*{1mm}$\mathbf{-\frac{V}{2}}$}
      \psfrag{n}[]{\raisebox{1mm}{\small\hspace*{5mm}$\mathbf{2e}$}}
      \includegraphics[scale=0.45]{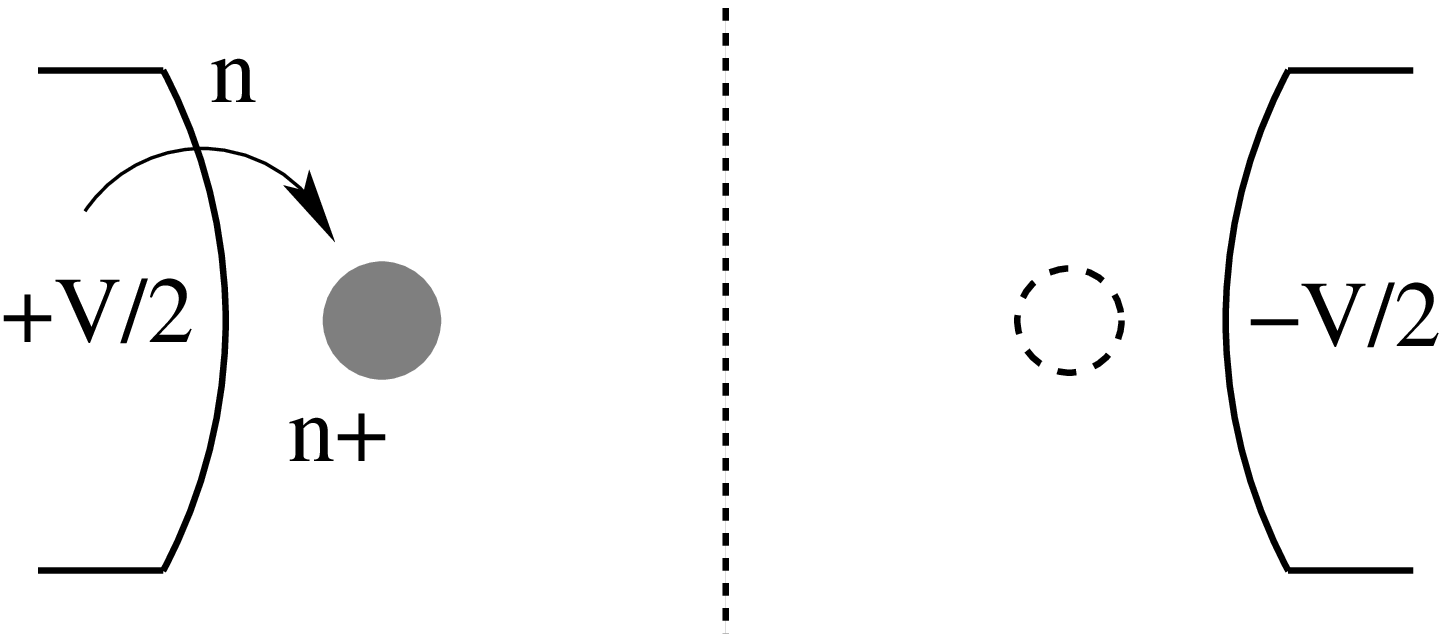}\\
      \begin{minipage}{0.4\textwidth}
        3: Being negatively charged, the grain is pushed back to the
        first lead, where the process starts over.
      \end{minipage}
    \end{center}
    \caption{Schematic illustration of the charge transfer mechanism in
      the shuttle regime. The grain performs oscillations back and
      forth between the leads, loading and unloading two charges at
      each turning point.}
    \label{fig:shut_reg_1}
  \end{figure}
  
  \newpage
  \begin{figure}[htb!]
    \begin{center}
      \psfrag{pos}{\hspace*{-3mm}\raisebox{-4mm}{\textbf{POSITIVE LEAD}}}
      \psfrag{neg}{\hspace*{-4mm}\textbf{NEGATIVE LEAD}}
      \psfrag{0.0}[]{}
      \psfrag{0.1}[]{\hspace*{2mm}\raisebox{-1mm}{\small{$\mathbf{0.1}$}}}
      \psfrag{0.2}[]{\hspace*{2mm}\raisebox{-1mm}{\small{$\mathbf{0.2}$}}}
      \psfrag{0.3}[]{\hspace*{2mm}\raisebox{-1mm}{\small{$\mathbf{0.3}$}}}
      \psfrag{0.4}[]{\hspace*{2mm}\raisebox{-1mm}{\small{$\mathbf{0.4}$}}}
      \psfrag{0.5}[]{\hspace*{2mm}\raisebox{-1mm}{\small{$\mathbf{0.5}$}}}
      \psfrag{0.6}{}
      \psfrag{t}{\small$\mathbf{t}$ $\mathbf{\mu}$s}
      \psfrag{x}{\hspace{-5mm}$\mathbf{X(t)/L}$}
      \psfrag{-}{\hspace{-2mm}$-\frac{1}{2}$}
      \psfrag{+}{\hspace{-2mm}$+\frac{1}{2}$}
      \includegraphics[scale=0.3]{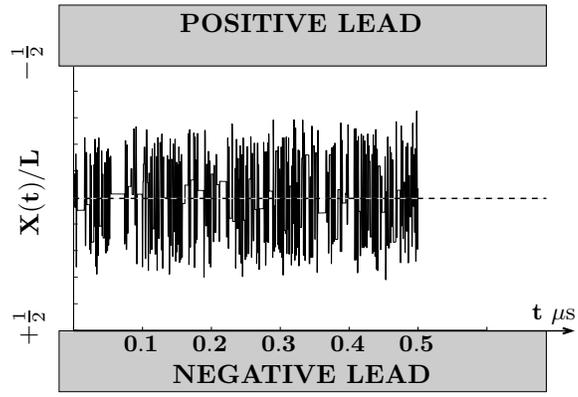}
      \vspace*{5mm}
      \caption{Plot of the position of the grain as a function of time
        for the bias voltage $V=2.0 V_0$ and the parameters $\alpha =
        6.4*10^{-4}$ and $\omega^2=4.27*10^{-3}$. For this voltage an
        uncharged grain is everywhere in tunneling contact with both
        leads so that the charge transfer cycle illustrated in
        \mbox{Fig. \ref{fig:shut_reg_1}} is possible.}
      \label{fig:xt_2}
    \end{center}
  \end{figure}

  \newpage
  \begin{figure}[htbp]
    \begin{center}
      \psfrag{+V/2}[]{\raisebox{4mm}
        {\hspace*{7mm}\Large $\mathbf{+\frac{V}{2}}$}}
      \psfrag{-V/2}[l]{\raisebox{4mm}{\Large $\mathbf{-\frac{V}{2}}$}}
      \psfrag{T}{\small$\mathbf{T}$}
      \psfrag{1}{\footnotesize{$\mathbf{1}$}}
      \psfrag{2}{\footnotesize{$\mathbf{2}$}}
      \psfrag{3}{\footnotesize{$\mathbf{3}$}}
      \psfrag{t}{$\mathbf{t}$}
      \includegraphics[scale=0.4]{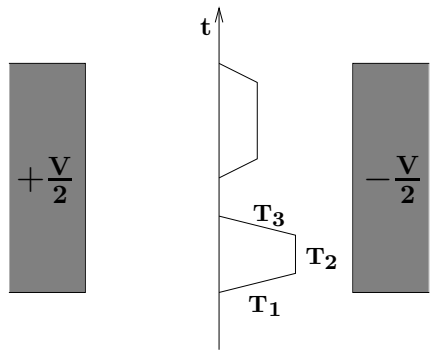}
      \vspace{5 mm}
      \caption{Schematic illustration of the three parts of the
        average half-period for a shuttle cycle discussed in the text.
        Two different kinds of period times are illustrated. The
        further the grain moves towards the lead, the shorter the time
        $T_2$ can be expected to be.}
      \label{fig:ill_1}
    \end{center}
  \end{figure}
  
  \newpage
  \begin{figure}[htb!]
    \begin{center}
      \psfrag{I (nA)}[b]{\hspace*{2mm}\raisebox{2mm}{\small$\mathbf{I (nA)}$}}
      \psfrag{V}[t]{\hspace*{2mm}\small$\mathbf{V/V_0}$}
      \includegraphics[scale=0.4]{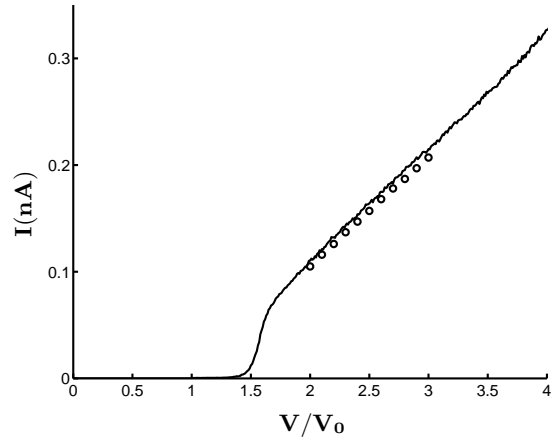}
      \vspace{5 mm}
      \caption{Comparison between the current obtained by Monte Carlo
        simulations of the system shown in \mbox{Fig.
          \ref{fig:system}} (solid line) and the current as calculated
        by using the analytical expression in \mbox{Eq.
          (\ref{eq:curr_1})} (circles). The parameters used in the
        simulation are: $\alpha = 6.4*10^{-4}$ and
        $\omega^2=4.27*10^{-3}$.}
      \label{fig:comp_2}
    \end{center}
  \end{figure}
  
\end{center}

\end{document}